# Using Smartphones as a Proxy for Forensic Evidence contained in Cloud Storage Services

George Grispos
University of Glasgow
g.grispos.1@research.gla.ac.uk

William Bradley Glisson
University of Glasgow
Brad.Glisson@glasgow.ac.uk

Tim Storer
University of Glasgow
timothy.storer@glasgow.ac.uk

**Abstract**

*Cloud storage services such as Dropbox, Box and SugarSync have been embraced by both individuals and organizations. This creates an environment that is potentially conducive to security breaches and malicious activities. The investigation of these cloud environments presents new challenges for the digital forensics community.*

*It is anticipated that smartphone devices will retain data from these storage services. Hence, this research presents a preliminary investigation into the residual artifacts created on an iOS and Android device that has accessed a cloud storage service. The contribution of this paper is twofold. First, it provides an initial assessment on the extent to which cloud storage data is stored on these client-side devices. This view acts as a proxy for data stored in the cloud. Secondly, it provides documentation on the artifacts that could be useful in a digital forensics investigation of cloud services[1].*

## 1. Introduction

Global connectivity, mobile device market penetration and use of remote data storage services are all increasing. Cisco reports that mobile data traffic reached 597 petabytes per month in 2011, which was over eight times greater than the amount of Internet traffic in 2000 [1]. They also predict that global mobile data transmission will exceed ten exabytes per month by 2016, with over 100 million smartphone users transmitting more than 1 gigabyte of data per month [1]. Supporting these predictions, cloud storage providers have experienced tremendous growth in the past year. A press release from Dropbox reported that their customer base has surpassed 25 million users [2]. They also claim that over one billion files are saved every three days using its services [3]. Box reports that enterprise revenue tripled in 2011 with mobile device implementation increasing 140% monthly [4]. Box have also experienced substantial penetration into the retail, financial and healthcare enterprise markets [5].

According to articles by CIO [6], surveys by Advanced Micro Devices (AMD) [7] and IBM [8], there is an apparent consensus that cloud computing is increasingly integrating into the business environment. The business reasons for this migration range from ideas like focusing on growth, innovation and customer value to improved use of resources, increasing employee productivity and cutting costs [8].

Cisco have argued that the storage and retrieval of corporate data from a number of cloud-based devices is concerning for security professionals, noting that "wherever users go, cybercriminals will follow" [9]. Support for this idea was demonstrated when Dropbox was used in corporate espionage [10]. Hence, there is no practical barrier to further utilization of cloud storage services to store and exchange illicit material. Gartner [11] warns that "investigating inappropriate or illegal activity may be impossible in cloud computing".

---

Investigating cloud service providers has been a topic of great discussion, with a number of challenges being raised about conducting a digital forensics investigation in such an environment [12-14]. One of these challenges is the investigator's ability to identify and recover digital evidence from the cloud in a forensically sound manner [14]. A public cloud environment managed by a third-party is potentially more difficult to investigate than a private cloud.

The remote, distributed and virtualized nature of the cloud obstructs the conventional, offline, approach to forensic evidence acquisition, which assumes the investigator has the ability to isolate and preserve the physical storage device in a timely manner, before directly obtaining a copy of the evidence for analysis. Although it is, in principle, possible for the storage device to be obtained from the cloud service provider, this process may take a significant amount of time, or be obstructed by cross-border jurisdictional disputes. In addition, the distributed nature of cloud architectures may make the identification of a single storage device containing relevant data impractical, because the data is actually stored in many different physical locations. Obtaining all of these devices may be extremely time consuming and expensive for a cloud service provider; and potentially disruptive to their customers. This situation prompts research into the following hypotheses:

> **H1:** Smartphone devices present a partial view of the data held in cloud storage services, which can be used as a proxy for evidence held on the cloud storage service itself.
>
> **H2:** The manipulation of different cloud storage applications influences the results of data collection from a smartphone device.

To address the hypotheses, the following questions were proposed:

1. To what extent can data stored in a cloud storage provider be recovered in a forensically sound manner from a smartphone device that has accessed the service?
2. What meta-data concerning the use of a cloud storage service can be recovered from a smartphone device?
3. What features of the cloud software influence the ability to recover data stored in a cloud storage service from a smartphone device?
4. What effect does user activity (such as, selecting files for download, or marking files for deletion) have on the ability to recover data stored in a cloud storage service from a smartphone device.

The contribution of this paper is two-fold. First, it provides a proof of concept that end-devices can be used to provide a partial view of the evidence in a cloud forensics investigation. This contribution focuses on tools currently available to practitioners providing a novel approach to practical solutions for emerging problems in the cloud. Second, it contributes to the documentation and evidentiary discussion of the artifacts created from specific cloud storage applications on iOS and Android smartphones.

The paper is structured as follows: Section 2 discusses the challenges of conducting digital forensic investigations in a cloud environment as well as presenting an overview of smartphone forensics. Section 3 describes the experimental design undertaken to address the research questions. Section 4 reports the findings, and a discussion of the results. Finally, Section 5 draws conclusions from the work conducted and presents future work.

## 2. Related Work

A growing number of researchers have argued that cloud computing environments are intrinsically harder to investigate than conventional computer artifacts [12-16]. The term 'cloud forensics' is defined as a "cross discipline of cloud computing and digital forensics" [13]. Ruan, et al., [13] described cloud forensics as a subset of network forensics. However, this definition does not take into consideration the virtualization aspect of the cloud. Ruan, et al., [13] also note that an investigation involving the cloud would include technical, organizational and legal aspects.

Grispos, et al., [16] described how digital forensic models and techniques used for investigating computer systems could prove ineffective in a cloud computing environment. Furthermore, Grispos, et al., identified several challenges for forensic investigators including: creating adequate forensic images, the recovery of segregated evidence and large data storage management. Reilly, et al., [17] speculated that one potential benefit of cloud computing is that the data being investigated will be located in a central location, which will mean incidents can be investigated quicker. This is unlikely to be the case. The very nature of a cloud potentially means that even evidence related to individuals within the same organization could be segregated in different physical locations [18]. Taylor, et al., [14] suggest there is also the possibility that potential important evidence could be lost in a cloud. Registry entries in Microsoft Windows platforms, temporary files, and meta-data could all be lost if the user leaves the cloud [14].

To enable a forensic investigation to be conducted, evidence needs to be collected from the cloud. This is likely to pose a great challenge to forensic investigators [19]. Researchers have begun proposing methods of acquiring evidence from a variety of cloud providers and services [19, 20]. Delport, et al., [20] proposed the idea of isolating a cloud instance for further investigation and several methods were proposed. None of these methods were empirically validated nor is it clear how a forensic image of the instance under investigation is obtained from the proposed techniques.

Dykstra and Sherman [19] have evaluated the performance of current forensic tools (FTK Imager and Encase Enterprise) and used these to acquire evidence from Infrastructure-as-a-Service (IaaS) instances stored in the Amazon Elastic Compute Cloud (EC2) [19]. Although the methods proposed were empirically evaluated, they are limited in essence to only IaaS virtual instances where a remote acquisition agent can be loaded on the instance under investigation. A second issue with this approach is that the investigator must be in possession of Amazon EC2 key pairs used to connect to the instance. The purpose of the key pairs is to ensure that only the instance's owner has access to the instance [21]. These public/private keys are created by the owner when the instance is first created using the Amazon Web Services Management Console [22]. Unless the investigator can recover these keys, then the methods of acquisition as described by Dykstra and Sherman cannot be used.

Taylor, et al., [14] have extensively examined the legal issues surrounding cloud computing and comment that any evidence gathered from the cloud should be conducted within local laws and legislation. Phillips [23] discussed the issue of the challenge of keeping a chain of custody for such an investigation, and asked: "Where physically is that part of the cloud at any given time? And where was it when forensically acquired?". Phillips argues that the cloud is a dynamic paradigm and physically isolating it to conduct an investigation could be a daunting task for the investigator.

When multiple devices are used to access data in the cloud, these problems are exacerbated. In particular, smartphones are increasingly being used to access data stored in a cloud while the user is mobile. Smartphone devices are distinguishable from a traditional mobile phone by its superior processing capabilities, a larger storage capacity, as well as its ability to run complex operating systems and applications [24]. From an evidentiary perspective, the smartphone can be considered a treasure trove of forensic evidence. A recent study recovered more than 11,000 data artifacts from just 49, predominately low-end, devices [25]. As with a traditional mobile phone, the smartphone not only stores call logs, text messages and personal contacts, but it also has the ability to store web-browsing artifacts, email messages, instant messenger lots, GPS coordinates, as well as third-party application related data [26-28].

There are a number of tools that can be used to perform a data acquisition from a smartphone. Examples of these include Cellebrite's Universal Forensic Extraction Device (UFED) and its associated tool suites [29]; Micro Systemation's XRY tools [30]; The Mobile Internal Acquisition Tool [31]; Paraben's Device Seizure [32] and RAPI Tools [33]. The availability of tools for acquiring data from smartphone devices makes it possible to investigate the interactions of these devices with cloud storage providers, without accessing data in the cloud service directly. However, there is currently a lack of evidence as to the relationship between the data held in the cloud and that retained by a smartphone following an interaction.

## 3. Experimental Design

To support the hypothesis proposed in the introduction, the experiment was broken into stages. The six stages included: 1) preparing the smartphone device and installing the cloud application; 2) loading a data set to a cloud storage provider; 3) connect to the data through the application on the smartphone; 4) perform various file manipulations to the data set and smartphone device; 5) process the device using the Universal Forensic Extraction Device (UFED); and 6) use a number of forensic tools to extract the files and artifacts from the resulting memory dumps.

UFED and its associated application the 'Physical Analyzer' along with FTK Imager are used in this experiment. The devices were processed with the UFED tools. The memory card used in the HTC Desire was processed using FTK. These tools were chosen based on practicality and availability to the authors.

Two smartphone devices were selected for use in this experiment: an Apple iPhone 3G and a HTC Desire (hereafter referred to as devices). Table 1 – Smartphone device features, highlights the notable features of these devices. These devices were selected for two reasons. First, they are compatible with the choice of forensic toolkit (UFED) used to perform a physical dump of the internal memory. Second, the operating systems used on these devices represent the two most popular smartphone operating systems in use [34]. A number of smartphone devices fulfill both these criteria and could have been used in the experiment. The decision to use these specific devices was a pragmatic decision based on device availability.

| Feature | iPhone 3G | HTC Desire |
| --- | --- | --- |
| Operating system | iOS v. 3 | Android v. 2.1 (Eclair) |
| Internal memory | 8 GB storage | 576 MB RAM |
| Memory card | No | Yes (4 GB) |

**Table 1: Smartphone device features**

The selection criteria for the smartphone devices limited the number of cloud storage applications available to only the applications compatible with both operating systems. The scope of the experiment was limited in the following ways:

- This experiment was conducted in the United Kingdom, where Global System of Mobile (GSM) is the predominant mobile phone type, therefore non-GSM mobile devices were not considered;

- A number of smartphone devices which run either iOS or Android were not considered due to compatibility issues with the toolkit, or were a pragmatic decision based on device availability; and

- Various cloud storage applications were not considered because they do not support either or both of the chosen operating system platforms.

The cloud storage applications selected for inclusion in this experiment are: Dropbox (iOS version 1.4.7, Android version 2.1.3), Box (iOS version 2.7.1, Android version 1.6.7) and SugarSync (iOS version 3.0, Android version 3.6).

A pre-defined data set was created which comprised 20 files made up of audio (mp3), video (mp4), image (jpg) and document (docx and pdf) file types. Table 2 – Data set, defines the files in this data set. The following steps were undertaken to prepare both devices for the experiment and were repeated every time the experiment was reset for a different cloud storage application.

1. The smartphone was 'hard reset', which involved restoring the default factory settings on the device. In the case of the HTC Desire, the SD memory card was forensically wiped using The Department of Defense Computer Forensics Lab tool dcfldd. These steps were done to remove any previous data stored on the devices and the memory card.
2. The device was then connected to a wireless network which was used to gain access to the Internet. The cloud storage application was downloaded and installed either via the Android or Apple 'app market', depending on the device used. The default installation and security parameters were used during the installation of the application.
3. The cloud storage application was executed, and a new user account was created using a predefined email address and a common password for that cloud storage application.
4. After the test account was created, the application was 'connected' to the cloud storage provider's services, which meant the device was now ready to receive the data set.
5. A personal computer running Windows 7 was used to access the test account created in Step 4 and the data set was then uploaded to the cloud storage provider using a web browser. The smartphone was synced with the cloud storage provider, to ensure the data set was visible via the smartphone application.

**Table 2. Data set**

| Filename | Size (bytes) | Manipulation |
|---|---|---|
| 01.jpg | 43183 | FV |
| 02.jpg | 6265 | FOF |
| 03.jpg | 102448 | NP |
| 04.jpg | 5548 | FVD |
| 05.mp3 | 3997696 | FV |
| 06.mp3 | 2703360 | FOF |
| 07.mp3 | 3512009 | NP |
| 08.mp3 | 4266779 | FVD |
| 09.mp4 | 831687 | FV |
| 10.mp4 | 245779 | FOF |
| 11.mp4 | 11986533 | NP |
| 12.mp4 | 21258947 | FVD |
| 13.pdf | 1695706 | FV |
| 14.pdf | 471999 | FOF |
| 15.pdf | 2371383 | NP |
| 16.pdf | 1688736 | FVD |
| 17.docx | 84272 | FV |
| 18.docx | 85091 | FOF |
| 19.docx | 14860 | NP |
| 20.docx | 20994 | FVD |

*Key: FV = File viewed, FOF = File viewed and saved for offline accessed, NP = no manipulation, FVD = file viewed and then deleted.*

6. When the entire data set was visible on the smartphone, a number of manipulations were made to files in the data set. Table 2 – Data set summarizes these manipulations. These included:
    - a file being viewed or played;
    - a file viewed or played then saved for offline access;
    - a file viewed or played then deleted from the cloud storage provider; and

- some files were neither opened/played nor deleted (no manipulation).

7. The smartphone and cloud storage application was also manipulated in one of the following ways:
    - *Active power state* - the smartphone was not powered down and the application's cache was not cleared;
    - *Cache cleared* - the applications cache was cleared;
    - *Powered off* - the smartphone was powered down; and
    - *Cache cleared and powered off* - the applications' cache was cleared and the smartphone was powered off.

   These were done to mimic various scenarios a forensic investigator could encounter during an investigation. The smartphone was then removed from the wireless network to prevent any accidental modification to the data set.

8. Immediately after the above manipulations, the smartphone device was processed to create a forensic dump of its internal memory. In the case of the HTC Desire, the Secure Digital (SD) memory card was also processed. The iPhone device was processed using the Physical Analyzer 'add-on', which is used to extract memory dumps of such devices. A step-by-step wizard provided instructions on how to prepare the device for the extraction. A number of forensic programs were then loaded onto the smartphone [35]. From the selection menu, the User partition was selected for extraction from the device, and the resulting memory dump was saved to a forensically wiped 16 GB USB flash drive. The entire process took approximately 1 hour and 10 minutes. On the USB flash drive a 7.56 GB forensic image was created, as well as a ufd file, which is used to load the forensic dump into the Physical Analyzer application for further analysis.

   The extraction process for the HTC Desire differed from that of the iPhone as this device was processed directly with the UFED. The SD card was processed separately from the smartphone using FTK Imager. A forensic image of the SD card was created using the default parameters for this application, and saved to a forensically wiped 16 GB USB flash drive. This process took approximately fifteen minutes and as a result a 3.8 GB image was created. Before the HTC Desire was processed using the UFED, the USB debugging option was enabled on the smartphone. This is required by the UFED to create the memory dumps from the device. The default parameters for a Physical Extraction on the UFED were selected, and the make and model of the device were provided. A forensically wiped 16 GB USB flash drive was used to save the resulting memory dumps. The process took approximately four hours to complete. As a result, six binary images were created on the USB flash drive, one for each partition on the device, as well as a ufd file.

9. The images extracted from the smartphone device were then loaded into Physical Analyzer, where the iOS and Android file systems were reconstructed. FTK 4 was used as the primary tool for analysis. This involved extracting the partitions from the dumps in Physical Analyzer and then examining them using FTK. Analysis techniques used included: string searching for the password, filtering by file types and browsing the iOS and Android file systems.

## 4. Analysis and Findings

A summary of the results of which files were recovered from the HTC Desire and iPhone devices is shown in Tables 3 – HTC Desire Files Recovered and 4 – iPhone Files Recovered. Several observations can be drawn from the results. Smartphone devices can be used to recover evidence from cloud storage services. The artifacts recovered include data stored in the storage service. This is provided that the user of the device has accessed these files in some way using the cloud storage provider's application. Files which were not viewed on the mobile phones were not recovered. The exception was the Joint

Photographic Experts Group (JPEG) image, where a thumbnail of the image was recovered. The results also show that the recovery of mp3 and mp4 files was not very successful.

**Table 3. HTC Desire files recovered**

| Filename | DB-AP | DB-CC | DB-PD | DB-CPD | B-AP | B-CC | B-PD | B-CPD | S-AP | S-CC | S-PD | S-CPD |
|---|---|---|---|---|---|---|---|---|---|---|---|---|
| *01* | T | T | T | T | ✓ | D | ✓ | D | ✓ | D | ✓ | D |
| *02* | ✓ | ✓ | ✓ | ✓ | ✓ | D | ✓ | D | ✓ | ✓ | ✓ | ✓ |
| *03* | T | T | T | T | T | T | T | T | T | T | T | T |
| *04* | T | T | T | T | ✓ | D | ✓ | D | ✓ | D | ✓ | D |
| *05* | | | | | ✓ | D | ✓ | D | | | | |
| *06* | ✓ | ✓ | ✓ | ✓ | ✓ | D | ✓ | D | ✓ | ✓ | ✓ | ✓ |
| *07* | | | | | | | | | | | | |
| *08* | | | | | ✓ | D | ✓ | D | | | | |
| *09* | | | | | ✓ | D | ✓ | D | | | | |
| *10* | ✓ | ✓ | ✓ | ✓ | ✓ | D | ✓ | D | ✓ | ✓ | ✓ | ✓ |
| *11* | | | | | | | | | | | | |
| *12* | | | | | ✓ | D | ✓ | D | | | | |
| *13* | ✓ | D | ✓ | D | ✓ | D | ✓ | D | ✓ | ✓ | ✓ | ✓ |
| *14* | ✓ | ✓ | ✓ | ✓ | ✓ | D | ✓ | D | ✓ | ✓ | ✓ | ✓ |
| *15* | | | | | | | | | | | | |
| *16* | D | D | D | D | ✓ | D | ✓ | D | ✓ | ✓ | ✓ | ✓ |
| *17* | ✓ | D | ✓ | D | ✓ | D | ✓ | D | ✓ | D | ✓ | D |
| *18* | ✓ | ✓ | ✓ | ✓ | ✓ | D | ✓ | D | ✓ | ✓ | ✓ | ✓ |
| *19* | | | | | | | | | | | | |
| *20* | D | D | D | D | ✓ | D | ✓ | D | ✓ | D | ✓ | D |

***Key: DB = Dropbox; B = Box; S = SugarSync; AP = Active Power state; CC = Cache cleared; PD = Powered Down; CPD = Cache Cleared and Powered Down; ✓ = File Found; T = Only Thumbnail Found; D = Deleted File Recovered.***

All the files marked for offline access were recovered from both devices. The SD memory card used in the Android device contained files which were either deleted by the user or deleted as a result of the application cache being cleared. Android-based devices allow files to be stored on either the device itself or on the memory card [36]. Clearing the application's cache has an adverse effect on the recovery of files. This is more evident on the iPhone, which does not contain an SD card. Powering down the smartphone devices did not have an effect on the recovery of data. As a result, the files recovered were identical to that of the active state scenario.

A variety of meta-data can be recovered from both devices. Meta-data recovered from the devices included email addresses used to register for the service, databases related to data stored in the service and transaction logs related to user activity. Apart from the Dropbox application on the iPhone, all other meta-data can still be recovered after the cache is cleared.

### 4.1. Android Applications

An analysis of the Android memory dumps revealed that forensic artifacts can be recovered from both the smartphone itself and the SD memory card. Files and meta-data related to the applications stored in internal memory can be found in a subfolder under the location */data/data/* [26]. Hoog discusses the

Android file system in further detail [26]. The location of evidence on the SD card varies depending on the application being investigated. Application data stored in internal memory is generally saved in a subfolder under the location

### 4.1.1. Dropbox.

Dropbox related artifacts can be recovered from two main locations on the SD card. JPEG image thumbnails can be found under */Android/data/com.dropbox.android/cache/thumbs/*. Files saved for offline viewing and documents viewed and not deleted on the device, can be found under the location */Android/data/com.dropbox.android/files/scratch*. An analysis of the 'unallocated space' revealed that the two document files which were deleted (16.pdf and 20.docx) were still physically stored on the SD card. These files were recovered by FTK. On the device itself, two SQLite databases contain the majority of meta-data evidence. These databases are found in the directory */data/data/com.dropbox.android/databases/*. The first database, *db.db* contains meta-data related to the files currently stored in the service, i.e., files which have not been deleted. This information is stored in a table called '*dropbox*'. Examples of the information that can be recovered include the file names, their size in bytes, offline storage information and location, the last modified time, and if the file has been designated as a 'favorite'.

The second database, *prefs.db*, contains user-specific meta-data such as the user's name and email address. This information is stored in a table called *DropboxAccountPrefs*. Clearing the cache of the Dropbox application, removes the documents viewed and not deleted on the device which are stored in the *com.dropbox.android/files/scratch* directory. These files were still physically stored on the SD card and recovered by FTK. The files saved for offline access, JPEG thumbnails and the contents of the SQLite databases remain unaltered.

### 4.1.2. Box.

Files stored in the Box storage service can be recovered from three locations on the SD card. The files saved for offline access can be recovered from the directory */Box/email_address/*, where *email_ address* is the email address used to register for the service. The Box application creates a cache of all the files the user has viewed on the smartphone. These can be recovered from the directory */Android/data/com.box.android/cache/filecache*. Fifteen files from the data set were found in this directory. The files missing, are those files which are marked as 'no manipulation' in Table 2 – Data set. A subdirectory of the above location called */tempfiles/box_tmp_images* contains thumbnails of the four JPEG images from the data set.

Meta-data artifacts about the Box service can be recovered from the smartphone. The Box application creates a JavaScript Object Notation (JSON) file called *json_static_model_emailaddress_0*, this is the email address used at registration. This file can be found in the directory */data/data/com.box.android/files/* and it contains meta-data related to the files stored in this Box account. The information which can be recovered includes: the name for each file stored in the account; the size of each file in bytes; the Secure Hash Algorithm (SHA) hash value of the file; a unique ID number for each file; a 'flag' indicating if the file has been shared with other individuals; and a UNIX timestamp of the date and time the last modification was made to the file. A second directory containing meta-data about the application can be found under */data/data/com.box.android/shared_prefs*. The Box service creates the following two XML files:

- *myPreference.xml* - which contains the authentication token associated with this particular account and the email address used to register for the Box service; and
- *Downloaded_Files.xml* - contains data about files downloaded to the SD card from the Box service. 'Long name' is the ID number assigned to that particular file and 'value' is the date and time the file was deleted and is stored as a UNIX timestamp.

When the cache is cleared on the Box application, the contents of the *Android/data/com.box.android/cache/filecache* and the */Box/email_address/* directories are deleted. All other files and meta-data related to the Box service are not affected.

### 4.1.3. SugarSync.

Files can be recovered from three main locations on the SD card. A directory called */.sugarsync*, contains the three PDF files viewed on the smartphone. A sub-directory of the above location called */.httpfilecache*, contains, the three JPEG images viewed on the device, thumbnails of the four images files, and four document files viewed (13.pdf, 16.pdf, 17.docx and 20.docx) on the device. The final location is a directory called */MySugarSyncFolders*, which contains the files saved for offline viewing. Meta-data related to the SugarSync service can be found on the smartphone device. A transactional log, which lists all the activities related to the service, can be found in the location */data/data/com.sharpcast.sugarsync/app_SugarSync/SugarSync/log/sugarsync.log*. Another source of meta-data related to the service is an SQLite database called *SugarSyncDB* which is stored in */com.sharpcast.sugarsync/databases*. In this database is a table called *rec_to_offline_ file_X*, X is a value appended to the end of the file. This table contains meta-data related to the files saved for offline access. This includes the name of the file saved offline, and a UNIX timestamp of when the file was saved. When the SugarSync cache is cleared the only files affected are those stored under the folder */.sugarsync/.httpfilecache*, which are deleted from the SD card.

## 4.2. iOS Applications

As the iPhone does not have an SD memory card, all the forensic artifacts recovered were from the internal memory. No deleted files were recovered. Files and meta-data related to the applications are stored under */private/var/mobile/Applications* in the User Partition [27]. Each application creates a directory under this path. The following discussion concerns evidence found under this location. Hoog and Strzempka [37], as well as Morrissey discuss the iOS file system in further detail [28].

### 4.2.1. Dropbox.

Dropbox creates a directory called */8D9651F8-C932-42AC-B3C0-9AF1BE5A1647*. Files can be recovered in the subfolder *Library/Caches/Dropbox*. Under this location, the following files were recovered:

- thumbnails images of the three JPEG images viewed on the device;
- the five files saved for offline access; and
- the pdf and docx files viewed but not deleted from the device.

No other files were found on the device. The meta-data depository is an SQLite database called *Dropbox.sqlite* which is found under a subfolder called */Documents/*. The *ZCACHEDFILE* table found in this database contains meta-data such as: the file names; the size of the files in bytes; the date and time the file was last viewed, stored as a MAC timestamp; a flag to indicate if the particular file is a 'favorite' or not, if the file is a favorite this flag is set to '1' else the flag is set to '0'; an ID number assigned to each file in the user's account; and if a thumbnail exists for that particular file.

A property list (plist) file stored under the subfolder */Library/Preferences/com.getdropbox.Dropbox.plist* contains the email address used for the Dropbox account and information related to files which were downloaded. A transaction log called *analytics.log* related to the user activity can be found under the subfolder */Library/Caches*. The timestamps used in this log are in the UNIX timestamp format and not the MAC format as used in the database discussed above. When the Dropbox cache is cleared, the only files remaining on the device are those saved for offline access. The *Dropbox.sqlite* database is also affected when the cache is cleared. This now contains only meta-data related to the five remaining files on the device. All other transaction logs remain unchanged.

**Table 4. iPhone files recovered**

| Filename | DB-AP | DB-CC | DB-PD | DB-CPD | B-AP | B-CC | B-PD | B-CPD | S-AP | S-CC | S-PD | S-CPD |
|---|---|---|---|---|---|---|---|---|---|---|---|---|
| 01 | T |   | T |   | T | T | T | T | ✓ |   | ✓ |   |
| 02 | ✓ | ✓ | ✓ | ✓ | ✓ | ✓ | ✓ | ✓ | ✓ | ✓ | ✓ | ✓ |
| 03 | T |   | T |   |   |   |   |   |   |   |   |   |
| 04 | T |   | T |   | T | T | T | T | ✓ |   | ✓ |   |
| 05 |   |   |   |   |   |   |   |   | ✓ | ✓ | ✓ | ✓ |
| 06 | ✓ | ✓ | ✓ | ✓ | ✓ | ✓ | ✓ | ✓ | ✓ | ✓ | ✓ | ✓ |
| 07 |   |   |   |   |   |   |   |   |   |   |   |   |
| 08 |   |   |   |   |   |   |   |   | ✓ | ✓ | ✓ | ✓ |
| 09 |   |   |   |   |   |   |   |   | ✓ |   | ✓ |   |
| 10 | ✓ | ✓ | ✓ | ✓ | ✓ | ✓ | ✓ | ✓ | ✓ | ✓ | ✓ | ✓ |
| 11 |   |   |   |   |   |   |   |   |   |   |   |   |
| 12 |   |   |   |   |   |   |   |   | ✓ |   | ✓ |   |
| 13 | ✓ |   | ✓ |   |   |   |   |   | ✓ |   | ✓ |   |
| 14 | ✓ | ✓ | ✓ | ✓ | ✓ | ✓ | ✓ | ✓ | ✓ | ✓ | ✓ | ✓ |
| 15 |   |   |   |   |   |   |   |   |   |   |   |   |
| 16 |   |   |   |   |   |   |   |   | ✓ |   | ✓ |   |
| 17 | ✓ |   | ✓ |   |   |   |   |   | ✓ |   | ✓ |   |
| 18 | ✓ | ✓ | ✓ | ✓ | ✓ | ✓ | ✓ | ✓ | ✓ | ✓ | ✓ | ✓ |
| 19 |   |   |   |   |   |   |   |   |   |   |   |   |
| 20 |   |   |   |   |   |   |   |   | ✓ |   | ✓ |   |

*Key: DB = Dropbox; B = Box; S = SugarSync; AP = Active Power state; CC = Cache cleared; PD = Powered Down; CPD = Cache Cleared and Powered Down; ✓ = File Found; T = Only Thumbnail Found; D = Deleted File Recovered.*

### 4.2.2. Box.

This application creates a directory called */6AF4F431-3CD2-477A-BD59-284782A0166F*. Data can be found in three locations under this directory. The files saved for offline access can be found in the subfolder */Documents/SavedFiles*. The thumbnails of the three JPEG images which were not deleted from the Box service, can be found in the subfolder */Library/Caches/Thumbnails*. No other files from the data set were recovered from Box. The final location of interest is an SQLite database called *BoxCoreData Store.sqlite* found under the subfolder */Documents/*. This database contains a table called *ZBOXBASE-COREDATA*, which stores meta-data related to the user account and files used in the Box service. Examples of this meta-data include:

- the username and email address used to create the Box account;
- a unique authentication token assigned to the user account,
- the name and the size of each file in bytes;
- information whether the file is shared or a 'favourite'; and
- a unique ID number assigned to the file.

Clearing the cache of the Box application has no effect on the data or metadata held on the device.

### 4.2.3. SugarSync.

The SugarSync application created a directory called */8E8333F9-F034-4B63-BD86-29A43946CC13*. Files and meta-data were recovered from various subfolders under this location. The cache for the SugarSync application was located in two subfolders. The three mp3 files viewed were recovered from the location */tmp/cache*. The jpeg, mp4, docx and pdf files viewed on the device were found under */tmp/http_cache*. The files saved for offline access can also be recovered from a folder called */MyIphone* which is located under the */Documents* directory. Meta-data related to the SugarSync application can also be found in the */Documents* directory. Account specific information can be recovered from a file called *ringo.appdata*. An SQLite database in the same directory called *Ringo.sqlite*, contains a table called *ZSYNCOBJECT*. This table can be used to recover meta-data related to the files saved for offline access. When the SugarSync application cache is cleared, the contents of the */http_cache* folder are deleted. All other forensic artifacts are not affected by clearing the cache.

### 4.3. Analysis Summary

The results described above can be used to provide answers to the research questions proposed in Section One. First, using mobile forensic toolkits, data can be recovered from a smartphone device which has accessed a cloud storage service. The results from the experiment have shown that it is possible to recover files from the Dropbox, Box and SugarSync services using smartphone devices. On the HTC Desire, both deleted and available files were recovered. The forensic toolkits recovered nine files from Dropbox, fifteen from Box and eleven from SugarSync. On the iPhone, depending on application and device manipulation either five or seven files were recovered from Dropbox, seven or fifteen from SugarSync and five from Box. No deleted application files were recovered from the iPhone. Certain files types were recovered more than other file types. For example, the results show that JPEG images produced thumbnails on the device, and very few mp3 and mp4 files were present on either device. It is also interesting to note that more deleted files were recovered from Box than Dropbox or SugarSync on the HTC Desire; while this pattern did not hold true for the iPhone. Second, meta-data was recovered from all the applications on both devices. This meta-data included transactional logs containing user activity, meta-data related to the files in the storage service and information about the user of the application. Third, the recovery of files from the smartphone device is impacted by the application's cache being cleared. The Box application on the iPhone was the only application where there was no difference in the number of files recovered from the active power state and the cache cleared state. The results show that when the cache was cleared in all the other instances, fewer files were recovered from this state. Finally, user actions on specific files have shown to influence the recovery of these files. For example, if a file has been viewed using the smartphone, there is the opportunity for it to be recovered using forensic toolkits. This is provided that the user has not deleted the file, or cleared the application's cache. Files saved for offline access by the user have been recovered from all the applications. For the Box application on the HTC Desire, these files were deleted when the cache was cleared. The recovery of these files is dependent upon them not being overwritten by new data on the external memory card.

Furthermore, the above discussion and the results from the experiment can be used to support the hypotheses proposed in Section One. H1, the smartphone devices in this experiment contain a partial view of the data held in the cloud storage service. This statement continues to hold when the device is powered down. Therefore, a smartphone device potentially presents a forensic investigator with a proxy view of the evidence held in the cloud storage service. In support of H2, clearing the application's cache has an adverse effect on evidence collection. An interesting observation is that the same cloud application stores information in different locations on different operating systems.

Using artifacts recovered from the Box application, it is possible to download further files from the Box service. These can include files that have not been found on the device itself. This is possible through a URL which was created by the Box API. The URL requires three pieces of information:

- The authentication token recovered from the myPreference.xml file found in the location */data/data/com.box.android/shared_prefs*;
- The unique ID number, ZBOXID, which is the ID number assigned to each file stored in the service. This information can be recovered from the *json_static_model_emailaddress_0* file stored in the directory */data/data/com.box.android/files/*. The investigator requires the ID number for each file they wishes to download from the Box service; and
- The URL: *https://mobile-api.box.com/api/1.0/download/auth_token/zboxid*, where *auth_token* is the authentication token for the account and *zboxid* is the ID number of the file to be downloaded.

This information can be merged to reconstruct a URL, which will result in the file associated with the *ZBOXID* being downloaded. For example, the URL: https://mobile-api.box.com/api/1.0/download/u5es7xli4xejrh89kr6xu14tks6grjn3/2072716265 can be used to recover the JPEG image 01.jpg. The Android is not unique in containing this information. The data needed to reconstruct the URL can be recovered from the iPhone device. Relevant artifacts can be found in the *BoxCoreDataStore.sqlite* database in the directory */Documents/*. Privacy and legal discussions associated with this practice are out of scope for this paper.

## 5. Future Work and Conclusions

Conducting digital forensic investigations in cloud computing environments is an increasingly challenging and complex task. The interest in addressing cloud computing forensics is growing in both academia and industry. The diverse range of devices able to access services in a cloud environment and the attractiveness of the cloud infrastructure model to organizations will mean that the ability to conduct sound forensic investigations will be crucial in the future.

The results from this research have shown that smartphone devices which access cloud storage services can potentially contain a proxy view of the data stored in a cloud storage service. The recovery of data from these devices can in some scenarios provide access to further data stored in a cloud storage account. From the client perspective, it can potentially provide a partial view of the data without access to the data provider. The recovery of this evidence is dependent on two factors. First, the cloud storage application has been used to view the files in the cloud. Second, the user has not attempted to clear the cache of recently viewed files.

Future work can examine several key areas that include extending the smartphone hardware, operating systems, application datasets, examining other mobile devices and investigating scenarios that involve multiple devices. Future research also needs to consider cloud computing from a corporate investigation perspective.

Research needs to be conducted with a greater variety of smartphone devices, operating systems and cloud storage applications. The focus of this experiment is to evaluate the results from this paper on a larger scale. This will focus on the identification of storage and usage patterns and solutions that could be helpful in a forensics investigation.

Since multiple devices are being used to access cloud environments, the application of this research to other mobile devices such as tablets, iPads, and readers needs to be examined. This includes an examination of devices running various operating systems. An experiment is currently being undertaken to access residual artifacts from Gmail, Mozy, Ubuntu One and Evernote on end-devices connected to these services.

The very nature of the cloud environment encourages users to access data through multiple devices. Hence, multiple devices need to be examined from the perspective of a more complete dataset. This research will address information from the client-side and the provider-side. On the client-side, does the examination of multiple devices provide a more complete picture of the dataset that is stored in the cloud?

Can patterns of use and timelines be established from the residual artifacts that are left on these devices? From the provider, what is the most effective approach to forensically acquire data? Do the tools that are currently in the market satisfy market needs for conducting investigations?

Current work is examining the data leakage risk that cloud applications introduce to corporate environments. The idea is to identify the implications from a corporate policy perspective and determine if these applications introduce opportunities for data leakage from the organization. If so, what is the most effective way to minimize risk and maximize employee productivity? This work will propose a set of security measures for both cloud providers and smartphone users to mitigate the potential risk of data leakage.

## 6. Acknowledgements

This work was supported by the A.G. Leventis Foundation. Any opinions, findings, conclusions or recommendations expressed in this paper are those of the authors and do not reflect the views of the A.G. Leventis Foundation.